\documentclass[10pt,final,journal,twoside]{IEEEtran}
\usepackage[T1]{fontenc}
\usepackage[mathscr]{eucal}
\usepackage{ifpdf}
\usepackage{cite}
\usepackage{graphicx}
\usepackage[cmex10]{amsmath}
\usepackage{amssymb}
\usepackage{algorithmic}
\usepackage{units}
\usepackage{setspace}
\usepackage{algorithm}
\usepackage{array}
\usepackage[tight,footnotesize]{subfigure}
\usepackage{amsthm}
\usepackage{multirow}
\usepackage{enumerate}
\usepackage{color}

\newtheorem{proposition}{Proposition}

\DeclareMathOperator*{\argmax}{argmax}	
\newcommand*{\affmark}[1][*]{\textsuperscript{*}}
\renewcommand{\baselinestretch}{0.95}
\begin{document}
%------------------------------->Tilte
\title{Computation Offloading and Activation of Mobile Edge Computing Servers: A Minority Game}
%------------------------------->Author
\author{
\IEEEauthorblockN{Shermila Ranadheera\affmark[1], Setareh Maghsudi\affmark[1], and Ekram Hossain, \textit{IEEE Fellow}\\}
\thanks{S. Ranadheera and S. Maghsudi contributed equally to this work. S. Ranadheera and E. Hossain are with the Department of Electrical and Computer Engineering, University of Manitoba, Winnipeg, Canada (e-mail: ranadhes@myumanitoba.ca, Ekram.Hossain@umanitoba.ca). 
S. Maghsudi is with the Department of Electrical Engineering and Computer Science, Technical University of Berlin, Berlin, Germany (e-mail:maghsudi@tu-berlin.de). 
}
}
\maketitle
%------------------------------------------------------------------------------------------------------>Abstract
\begin{abstract}
With the ever-increasing popularity of resource-intensive mobile applications, Mobile Edge Computing (MEC), e.g., offloading computationally expensive tasks to the cellular edge, has become a prominent technology for the next generation wireless networks. Despite its great performance in terms of delay and energy, MEC suffers from restricted power allowance and computational capability of the edge nodes. Therefore, it is imperative to develop distributed mechanisms for computation offloading, so that not only the computational servers are utilized at their best capacity, but also the users' latency constraints are fulfilled. In this paper, by using the theory of Minority Games, we develop a novel distributed server activation mechanism for computational offloading. Our scheme guarantees energy-efficient activation of servers as well as satisfaction of users' quality-of-experience (QoE) requirements in terms of latency. 
\end{abstract}
%------------------------------------------------------------------------------------------------------>Keywords
\section{Introduction}
\label{intro}
Due to the ever-increasing popularity of computationally intensive applications, computation offloading capability has become a prerequisite for next generation wireless networks. Since the energy, storage, and computing capacity of small mobile devices are limited, mobile users need to transfer computationally expensive tasks to powerful computing servers. Despite its higher computational capability, remote cloud may not be the ideal option, as the long distance between the cloud and the user device yields substantial latency and energy cost. In contrast, small scale computing servers located in the network edge might provide services at reduced latency and energy cost, compared to the remote cloud. This is referred to as Mobile Edge Computing (MEC) \cite{8057148}. However, efficient utilization of MEC servers is vital, since they have limited computational resources and power. To this end, one solution is to activate only a specific number of servers, while keeping the rest in the energy saving mode. At the same time, users' latency requirements should be taken into account, as overloading the servers with computational tasks can result in unacceptable delay. Therefore, addressing this trade-off is a major issue in developing efficient MEC systems. This becomes challenging in the presence of uncertainty in task arrival and/or in the  absence of any central controller. Other challenges include minimizing users' energy consumption and efficient radio resource management. 

In \cite{8057148}, the authors develop a distributed algorithm in a game theoretic framework to address the  decision making problem for computation offloading by the users, so that the MEC cloud and radio resources are efficiently utilized. In \cite{7996854}, the authors investigate a computational offloading problem, where mobile users offload to a variety of edge nodes such as small base stations, macro base stations and access points, in order to utilize their computational resources. Reference \cite{7762913} provides centralized resource allocation algorithms that minimize the weighted sum energy consumption under delay constraints for both TDMA and OFDMA protocols in a mobile edge computation offloading system. A multi-objective offloading problem is formulated and analyzed using queuing theory in \cite{Li:80}. A comprehensive survey on the state-of-the-art of computation offloading in mobile edge networks can be found in \cite{7879258}. 

The majority of the existing literature focus on user-centric objectives such as meeting users' delay constraints and minimizing users' energy consumption. On the contrary, our work presents a hybrid view where both servers' and users' standpoints are considered. In doing so, we address the uncertainty caused by the randomness in channel quality and users' requests. We first analyze the statistical characteristics of the offloading delay. Based on this, we model the computational offloading problem as a planned market, where the price of computational services is determined by an authority. Afterward, by using the theory of minority games \cite{Challet:2014:MGI:2616207}, we develop a novel approach for efficient mode selection (or activation) at the servers' side. The designed mode selection mechanism guarantees a minimal server activation to ensure energy efficiency, while meeting the users' delay constraints. Moreover, this scheme is distributed, and does not require any prior information at the servers' side. We numerically investigate the performance of the proposed method.

In Section \ref{system_model}, we present the system model and formulate the server activation problem. In Section \ref{MGmodel}, we cast the servers' mode selection problem as a minority game theoretic framework, and provide an algorithmic solution. Numerical results and discussions are presented in Section \ref{simulation}. 
%------------------------------------------------------------------------------------------------------>System model
\section{System model and problem formulation}
\label{system_model}
We consider an MEC system consisting of a virtual pool of $M$ computational servers (e.g., small base stations), denoted by a set $\mathcal{M}$, and a set of users (e.g., mobile devices). Each user has some delay sensitive computational tasks to be completed in consecutive offloading periods. Each offloading period is referred to as one \textit{time slot}. In every time slot $t$, the users offload a total number of $K_{\textup{T}}$ computational jobs to the pool. Prior to task arrival, every server independently decides whether to 
\begin{itemize} 
\item accept computation jobs (\textit{active} mode); or 
\item not to accept any computation job (\textit{inactive} mode). 
\end{itemize} 
To become active, each server incurs a fixed energy cost represented by $e_{\textup{f}}$ (dimensionless value). In addition, doing 
\textit{each task} yields an extra $e_{\textup{j}}$ units of energy cost. By processing each job, a server receives a reimbursement 
(benefit) equal to $e_{\textup{p}}>e_{\textup{j}}$. Let $c(t)$ be the number of servers that decide to become active at time slot $t$. The total $K_{\textup{T}}$ jobs are equally divided among the active servers, so that the number of jobs per active server is given by
\begin{equation} 
\label{C_value} 
k(t) = K_{\textup{T}}/c(t). 
\end{equation}
Hence, each active server processes $k(t)$ jobs, and thus earns a total reward given by 
\begin{equation} 
\label{eq_Rth} 
R(t) = k(t)(e_{\textup{p}}-e_{\textup{j}})-e_{\textup{f}}. 
\end{equation} 
For each server, being in active mode is attractive only if a minimum desired reward, denoted by $R_{\textup{th}}>0$ can be obtained. Then each active server has to receive at least
\begin{equation} 
\label{eq_Ks} 
k_{\min} = \frac{R_{\textup{th}}+e_{\textup{f}}}{e_{\textup{p}}-e_{\textup{j}}} 
\end{equation} 
jobs to achieve the minimum desired reward. Each computational job requires a random time to be processed by a server, denoted by $t_c$. We assume that $t_c$ lies within the interval $(0,T)$ and has a truncated normal distribution with parameters $\mu$ and $\sigma$. Moreover, considering Rayleigh fading, the channel gain ($h$) is exponentially distributed with parameter $\nu$. We model the round trip transmission delay (from the user to the servers pool) as a linear function of the channel gain. The channel gains in both directions are assumed to be equal. Thus formally,\footnote{Assuming a normal distribution for the time required to perform each task does not limit the applicability, and similar analysis can be performed with any other distribution. The same holds for the linear model of the transmission delay. Any other model can be used at the expense of additional calculus steps. Also, note that since the required energy to perform each task is proportional to the required time to perform that task, one might consider $e_{\textup{j}} \propto \kappa \mu$ with $\kappa>0$.}
\begin{equation}
t_0=2(ah+b),
\end{equation} 
where $a<0$ and $b>0$ are constants. 
The total offloading delay $\theta$ is the sum of processing delay at the server, $t_c$ and the round trip transmission delay $t_0$. Thus, \begin{equation} 
\label{eq_theta} 
\theta=t_c+t_0.
\end{equation}
The following proposition characterizes $\theta$ statistically.\footnote{The proof follows by simple probability rules given the independence of $t_{c}$ and $t_{0}$. We omit the proof due to space limitation.} 
\begin{proposition} 
\label{pdftheta}
The expected value and variance of $\theta$ are given by
\begin{equation}
\mu_{\theta} = \mu + \sigma\frac{\phi(\frac{-\mu}{\sigma})-\phi(\frac{T-\mu}{\sigma})}{\Phi(\frac{T-\mu}{\sigma})-\Phi(\frac{-\mu}{\sigma})} +\frac{2(a+b\nu)}{\nu},
\end{equation} 
and
\begin{eqnarray}
\label{var_theta} \nonumber 
\sigma_{\theta}^2 &=& \sigma^{2}\Bigg(1+\left(\frac{\frac{-\mu}{\sigma}\phi(\frac{-\mu}{\sigma})-\frac{T-\mu}{\sigma}\phi(\frac{T-\mu}{\sigma})}{\Phi(\frac{T-\mu}{\sigma})-\Phi(\frac{-\mu}{\sigma})}\right)  \nonumber \\ 
&-&\left(\frac{\phi(\frac{-\mu}{\sigma})-\phi(\frac{T-\mu}{\sigma})}{\Phi(\frac{T-\mu}{\sigma})-\Phi(\frac{-\mu}{\sigma})}\right)^2 \Bigg) + \frac{4a^2}{\nu^2},
\end{eqnarray} 
respectively, where $\Omega=\frac{1}{\sigma\left(\Phi(\frac{T-\mu}{\sigma})-\Phi(\frac{-\mu}{\sigma})\right)}$. Moreover,
$\Phi(\epsilon)=\frac{1}{2}\left(1+\textup{erf}\left(\frac{\epsilon}{\sqrt{2}}\right)\right)$ is the cumulative distribution function of a standard normal distribution.
\end{proposition}
Every user requires its offloaded job(s) be completed by a deadline $T$. Therefore, in every round $t$ and for every server, the total processing time of all tasks, i.e., 
\begin{equation} 
\tau=\sum^{k(t)}_{i=1}\theta_i, 
\end{equation}
should be less than $T$, so that the delay experienced by the last user in the queue does not exceed the deadline $T$ as well. In other words, the condition $\tau \leq T$ ensures that all users receive their jobs completed before the deadline. Since $\theta_i$ are independent and identically distributed (i.i.d.), $\tau$ is the sum of $k(t)$ i.i.d. random variables. Therefore, the expected value and variance of 
$\tau$ are given by $k(t)\mu_{\theta}$ and $k(t)\sigma^{2}_{\theta}$, respectively. For large enough $k(t)$ (e.g., $k(t) \geq 30$), and by using the central limit theorem, the distribution of $\tau$ can be approximated as
\begin{equation} 
\label{eq:pdfoftau} 
\tau \sim \textup{Nor}(k(t)\mu_{\theta}, k(t)\sigma_{\theta}^2).
\end{equation}

Due to the uncertainty caused by the randomness, deterministic performance guarantee in terms of delay is not feasible. Thus we resort to a probabilistic guarantee of users' QoE requirement. Formally, let $\textup{Pr}[\tau>T] $ be the probability that $\tau$ exceeds $T$, i.e., the likelihood that the delay requirement of some offloading user(s) is not satisfied. We require that $\textup{Pr}[\tau>T]$ remains below a predefined threshold $\beta$. That is,
\begin{equation} 
\label{eq_qoe} 
\textup{Pr}[\tau>T] \leq \beta.
\end{equation}
Considering both the servers' and users' perspectives, the trade-off in the system can be seen as follows: On one hand, for each server it is beneficial to be active only if the number of active servers is less than a certain threshold $c_{\max}$, so that every active server receives the minimum number of jobs required to achieve the threshold reward (as stated by (\ref{eq_Ks})). On the other hand, the users prefer that the number of jobs per server $k(t)$ is small enough so that their desired QoE is fulfilled with high probability, i.e., the number of active servers shall be larger than a certain threshold $c_{\min}$. In what follows, we will derive the values of $c_{\min}$ and $c_{\max}$ analytically. Therefore, for the offloading system to perform efficiently, the number of active servers at any offloading round $t$, i.e., $c(t)$ should be determined in way that both servers and users are satisfied. We denote this value by $c_{\textup{th}}$.
%------------------------------------------>Sub
\subsection{Condition for Servers}
Recalling (\ref{eq_Ks}), to achieve minimum desired reward $R_{\textup{th}}$, each active server has to receive at least $k_{\min}$ jobs. Consequently, at most
\begin{equation} 
\label{cmax} 
c_{\max}=K_{\textup{T}}/k_{\min}
\end{equation}
servers can be in the active mode so that every active server receives the threshold reward $R_{\textup{th}}$, while inactive servers receive no reward. Thus, the condition below should be satisfied when selecting the cut-off $c_{\textup{th}}$:
\begin{equation} 
\label{servers}
\text{Condition I:}~~~~~~~~c_{\textup{th}}\leq c_{\max}. 
\end{equation}
%
%------------------------------------------>Sub
\subsection{Condition for Users}
Recall that the users' QoE requirement given by (\ref{eq_qoe}). Then, from (\ref{eq:pdfoftau}) and (\ref{eq_qoe}), we have
\begin{equation}
1-\Phi \left(\frac{T-k(t)\mu_{\theta}}{\sqrt{k(t)}\sigma_{\theta}}\right) \leq \beta, 
\end{equation} 
which, by definition, is equivalent to
\begin{equation} 
\label{eqerf} 
\textup{erf}\left(\frac{T-k(t)\mu_{\theta}}{\sqrt{2k(t)}\sigma_{\theta}}\right) \geq 1-2\beta. 
\end{equation}
Since $\textup{erf}(x)$ is an increasing function, $\textup{erf}^{-1}(x)$ is also an increasing function. Therefore, (\ref{eqerf}) results in
\begin{equation} 
k(t)\mu_{\theta}+\sqrt{2k(t)}\sigma_{\theta}\textup{erf}^{-1}(1-2\beta)-T \leq 0. 
\end{equation}
Solving the quadratic inequality, we obtain;
\begin{eqnarray} 
\label{K_value}
\frac{-\sqrt{2}\sigma_{\theta}\text{erf}^{-1}(1-2\beta)-\sqrt{\Delta}}{2\mu_{\theta}}\leq\sqrt{k(t)}\leq \nonumber \\ 
\frac{-\sqrt{2}\sigma_{\theta}\text{erf}^{-1}(1-2\beta)+\sqrt{\Delta}}{2\mu_{\theta}}
\end{eqnarray}
where $\Delta = 2\sigma_{\theta}^{2}\left(\text{erf}^{-1}(1-2\beta)\right)^2+4\mu_{\theta}T$. Since $k(t) \geq 0$, considering only the right hand side of the inequality (\ref{K_value}), we have 
\begin{equation} 
\label{K_users}
k(t)\leq\left(\frac{-\sqrt{2}\sigma_{\theta}\text{erf}^{-1}(1-2\beta)+\sqrt{\Delta}}{2\mu_{\theta}}\right)^2.
\end{equation}
Therefore,
\begin{equation} 
\label{Ku}
k_{\max}=\left(\frac{-\sqrt{2}\sigma_{\theta}\text{erf}^{-1}(1-2\beta)+\sqrt{\Delta}}{2\mu_{\theta}}\right)^2
\end{equation}
is the maximum allowable number of tasks per active server so that the users' QoE (i.e., latency) requirement is satisfied with probability $1-\beta$. Thus by (\ref{C_value}), the minimum number of active servers $c_{\min}$ to guarantee the users' QoE satisfaction is 
\begin{equation} 
\label{cmin} 
c_{\min}=\frac{K_{\textup{T}}}{k_{\max}}.
\end{equation} 
Therefore, the condition below should be satisfied when selecting the threshold $c_{\textup{th}}$.
\begin{equation} 
\label{users}
\textup{Condition II:}~~~~~~~~c_{\textup{th}} \geq c_{\min}.
\end{equation}
By conditions (\ref{servers}) and (\ref{users}), the optimal number of active servers, $c_{\textup{th}}$, is determined by solving the following equation:
\begin{equation} 
c_{\min} = c_{\max}. 
\end{equation}
Or equivalently, the system performs optimally in terms of servers' energy and users' delay when $c_{\textup{th}}$ servers are active so that 
\begin{equation} 
\label{eqCth} 
k_{\min}=k_{\max}. 
\end{equation} 
Thus, we obtain the threshold $c_{\textup{th}}$ using (\ref{cmax}), (\ref{cmin}), and (\ref{eqCth}) as
\begin{equation} 
\label{cth} 
c_{\textup{th}}=\frac{K_{\textup{T}}}{k_{\max}}. 
\end{equation}
To ensure that the entire system works efficiently, in addition to the optimal number of active servers ($c_{\textup{th}}$), the price of receiving computing services (i.e., $e_{\textup{p}}$) must be determined by an authority (for instance, macro base station or network 
planner). By using (\ref{eq_Ks}), we have
\begin{equation} 
\label{eqLp} 
e_{\textup{p}}=\frac{R_{\textup{th}}+e_{\textup{f}}}{k_{\max}}+e_{\textup{j}}, 
\end{equation}
with $k_{\max}$ given by (\ref{Ku}). In fact, in a distributed system, if a price larger than (\ref{eqLp}) is charged, more servers than 
$c_{\textup{th}}$ would become active, since every server achieves $R_{\textup{th}}$ with lower number of tasks than $k_{\min}$. In contrast, for $e_{\textup{p}}$ lower than (\ref{eqLp}), achieving $R_{\textup{th}}$ requires more tasks per server than $k_{\max}$, so that users' QoE might not be satisfied. 

Now the challenge is to activate $c_{\textup{th}}$ servers in a self-organized manner, which is addressed in the following section.    
%------------------------------------------------------------------------------------------------------>MG
\section{Modeling the problem as a Minority Game}
\label{MGmodel}
A Minority game (MG) can model the interaction among a large number of players competing for limited shared resources. In a basic MG, the players select between two alternatives and the players belonging to the {\em minority} group win. The minority is typically defined using some cut-off value. The collective sum of the selected actions by all players is referred to as the 
\textit{attendance}. The advantages of MG include simple implementation, low overhead, and scalability to large set of players, which are of vital importance in a dense wireless network. Details can be found in 
\cite{Challet:2014:MGI:2616207, 7891693}.

We model the formulated server mode selection problem as an MG, where the $M$  servers represent the players, with a cut-off value $c_{\textup{th}}$ for the number of active servers. In each offloading period, the servers decide between the two actions, i.e., being \textit{active} or \textit{inactive}, denoted by $1$ and $0$, respectively. We denote the action of a given player $i$ in the time slot $t$ by 
$a_i(t)$. The number of active servers $c(t)$ maps to the attendance. Each player has $S$ strategies. According to our formulated servers' mode selection problem and the analysis in Section \ref{system_model}, 
\begin{itemize}
\item If $c(t)\leq c_{\textup{th}}$, each of the $c(t)$ active servers (the minority) earns a reward higher than or equal to the 
      minimum desired reward, $R_{\textup{th}}$.
\item If $c(t)>c_{\textup{th}}$, $c(t)$ active servers cannot achieve $R_{\textup{th}}$. In this case inactivity (i.e., the action of 
      the minority) is considered as the winning choice, since inactive servers spend no cost without being properly reimbursed.
\end{itemize}
%-------------------------------->Sub
\subsection{Control Information}
After each round of play, a central unit (e.g., a macro base station) broadcasts the winning choice to all servers by sending a 
one-bit control information:
\begin{align}
w(t) = \begin{cases}
1, & \text{if}\ c(t) \leq c_{\textup{th}}\\
0, &  \text{otherwise.}
\end{cases}
\end{align}
Note that neither the actual attendance value $c(t)$ nor the system cut-off $c_{\textup{th}}$ is known by the players.
%-------------------------------->Sub
\subsection{Utility}
Let $U_{i,\textup{a}}(t)$ and $U_{i,\textup{p}}(t)$ denote the utility that server $i$ receives for being active and being inactive, respectively. Based on the discussion above, we define
\begin{align} 
\label{eq:utility_active}
U_{i,\textup{a}}(t) = \begin{cases}
1, & \text{if}\ c(t) \leq c_{\textup{th}} \\
0, & \text{otherwise}
\end{cases}
\end{align}
and 
\begin{align} 
\label{eq:utility_inactive}
U_{i,\textup{p}}(t) = \begin{cases}
1, & \text{if}\ c(t) > c_{\textup{th}} \\
0, & \text{otherwise.}
\end{cases}
\end{align}
%----------------------------------->Sub
\subsection{Distributed Learning Algorithm}
\label{algorithm}
Every player applies a basic strategy reinforcement technique to solve the formulated MG, summarized in \textbf{Algorithm \ref{Algo}} for some 
player $i$. Details can be found in \cite{Challet:2014:MGI:2616207}. 
%--------------->Alg
\begin{algorithm}
\caption{Distributed learning algorithm to solve server mode selection MG \cite{Challet:2014:MGI:2616207}}
\label{Algo}
\begin{algorithmic}[1]
\STATE \textbf{Initialization}: Randomly draw $S$ strategies from the universal strategy pool, gathered in a set $\mathcal{S}$. 
       Moreover, For every $s \in \mathcal{S}$, set the score $V_{i,s}(0)=0$.
\FOR{$t=1,2,...$}
  \STATE If $t=1$, select the current strategy, $s_{i}(1)$, uniformly at random from the set $\mathcal{S}$. Otherwise, select the best 
	       strategy so far, defined as 
  \begin{equation}
  \label{eq:best}
  s_{i}(t)=\argmax \limits_{s \in \mathcal{S}} V_{i,s}(t).
  \end{equation}
 \STATE Select the action $a_{i}(t)$, predicted by $s_{i}(t)$ as the winning choice.
 \STATE The central unit broadcasts the control information (winning choice), $w(t)$.
 \STATE Update the score of the strategy $s_{i}(t)$ as
 \begin{equation}
 \label{eq:Update}
 V_{i,s}(t+1)= \begin{cases}
 V_{i,s}(t)+1, & \text{if}\ a_{i}(t)=w(t) \\
 V_{i,s}(t), & \text{otherwise}
 \end{cases}
 \end{equation}
\ENDFOR
\end{algorithmic}
\end{algorithm}
%---------------->
%------------------------------------------------------------------------------------------------------->Section
\section{Numerical Results}
\label{simulation}
For numerical analysis, we choose $M=21$, $K_\textup{T}=500$, $R_{\textup{th}}=100$, $e_{\textup{j}}=5$, $e_{\textup{f}}=50$, 
$\beta=0.05$, $T=0.35$, $\mu=7$, $\sigma^2=2$, $\nu=1$, $a=-1$ and $b=2.5$. Simulation is carried out for $32$ runs and in each run, the servers randomly draw a set of strategies ($S=2$) and repeatedly execute the MG for $10000$ offloading periods. For the given parameters' value, using (\ref{cth}), the cut-off yields $c_{\textup{th}}=15$. The optimal (central activation) and random choice game (each server selects its action uniformly at random) are also simulated for comparison. 

\begin{figure}[!htb]
  \centering
	\includegraphics[width=0.32 \textwidth]{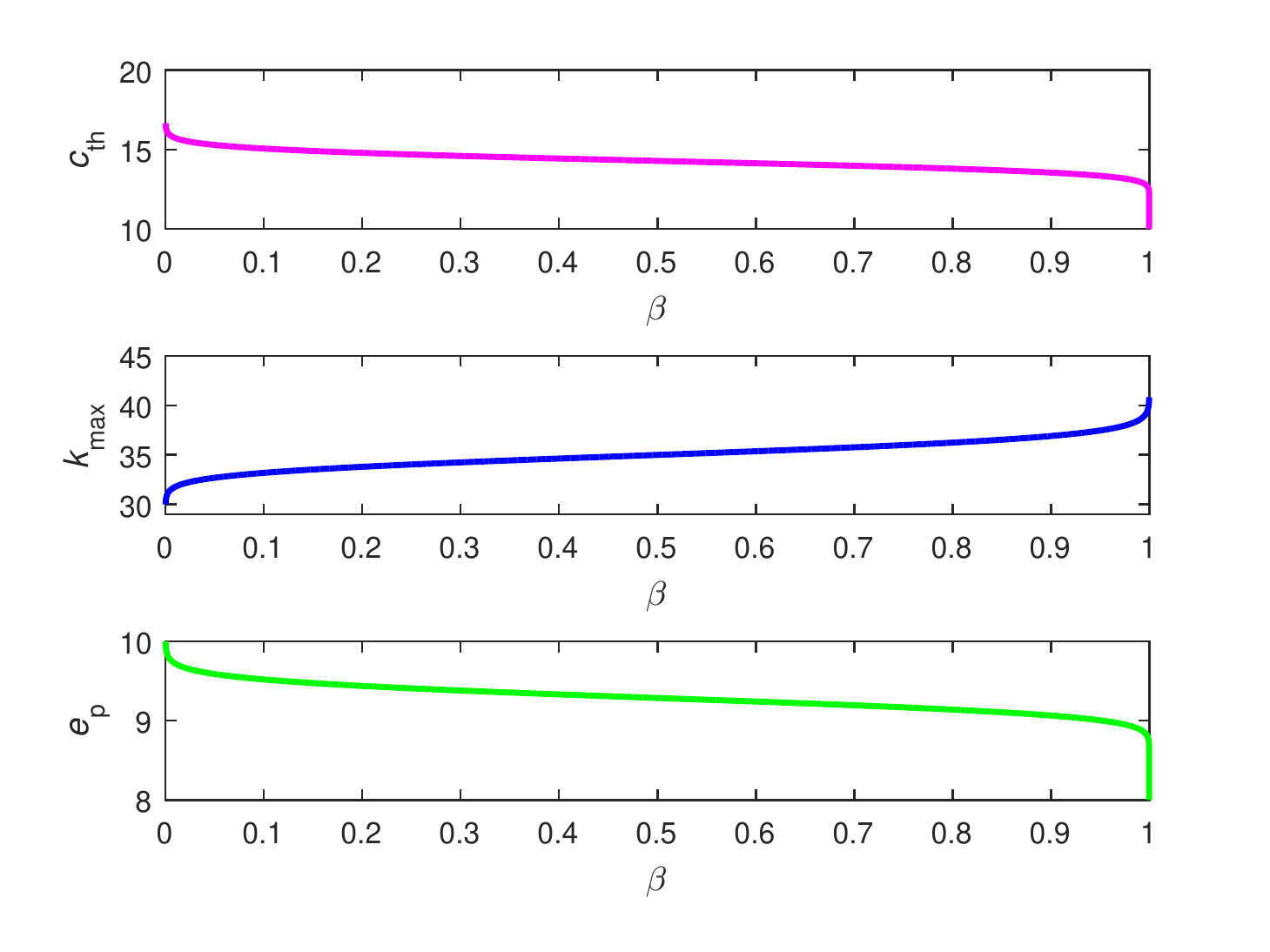}
  \caption{Variation of $c_{\textup{th}}$, $k_{\max}$, and $e_{\textup{p}}$ as a function of $\beta$.}
	\label{C_th_Ku_Lp_vs_beta}
\end{figure}

In Fig. \ref{C_th_Ku_Lp_vs_beta}, we present the variation of important system parameters as a function of users' QoE index $\beta$ 
(see Section \ref{system_model}). From the figure, the following can be concluded: As $\beta$ increases, the number of required active servers ($c_{\textup{th}}$) decreases, thereby allowing a larger number of offloading tasks to be processed per server. Similarly, the maximum allowable number of tasks per active server ($k_{\max}$) increases with increasing $\beta$. In fact, with larger $\beta$, larger delay ($T$) is tolerable, or in other words, longer task queue ($k_{\max}$) is allowed. Naturally, in this case, the price per task, 
$e_{\textup{p}}$, reduces, as intuitively expected for a weaker service.

\begin{figure}[!htb]
  \centering
	\includegraphics[width=0.32 \textwidth]{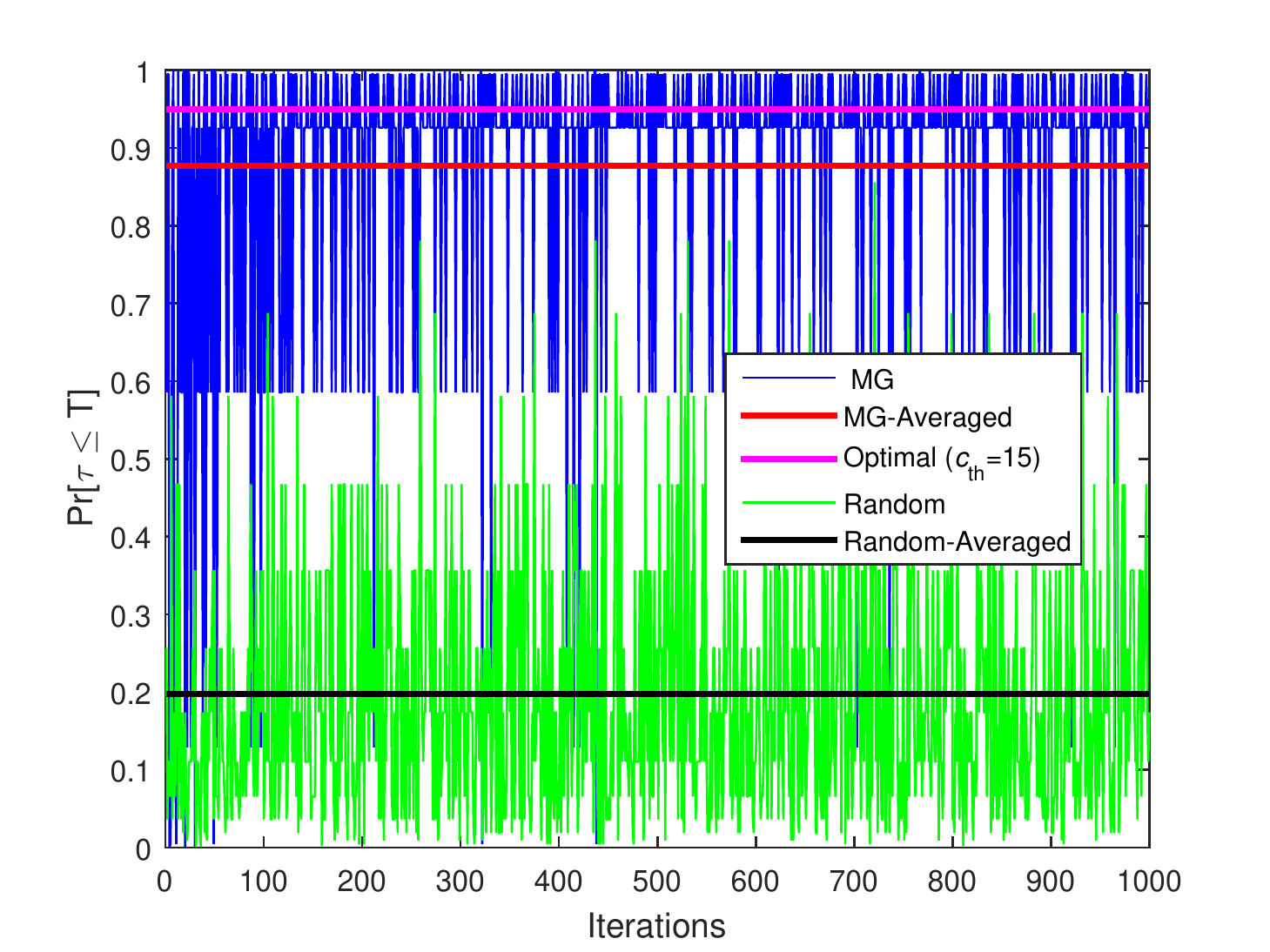}
   \caption{Users' QoE measure.}
	\label{prob_vs_time}
\end{figure}

\begin{figure}[!htb]
  \centering
	\includegraphics[width=0.32 \textwidth]{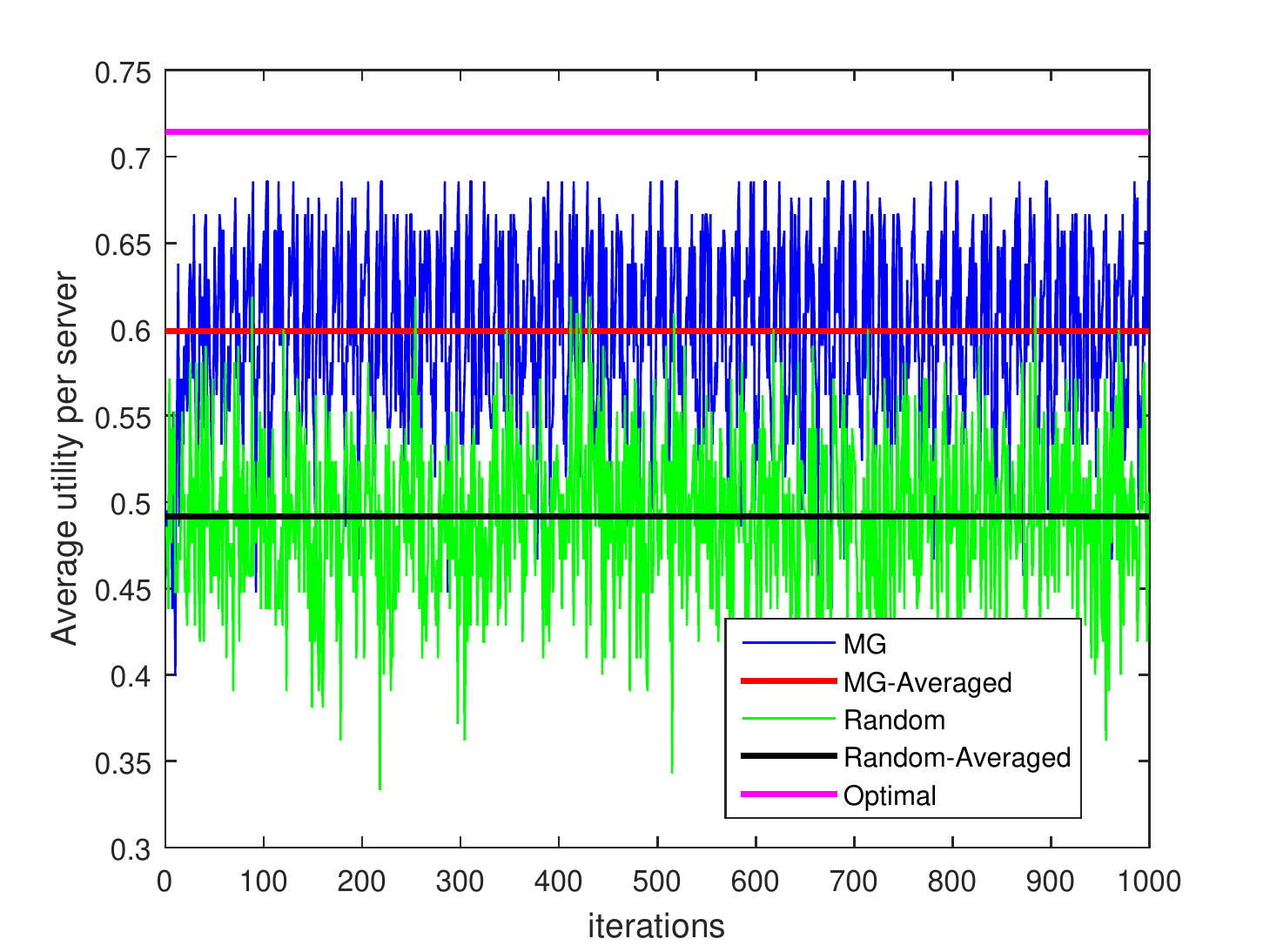}
   \caption{Average utility of a server.}
	\label{avg_utility}
\end{figure}

Fig. \ref{prob_vs_time} shows the changes in users' probability measure, i.e., $\text{Pr}[\tau>T]$. The users meet their QoE certainty requirement whenever $c(t)>c_{\textup{th}}$. As the attendance fluctuates near $c_{\textup{th}}$, the probability value also remains near the desired certainty.
The average utility per user is depicted in Fig. \ref{avg_utility}. It can be seen that the utility of MG-based strategy is higher than that of random selection. Yet, it is below the average utility of the optimal scenario. This is due to the fact that in MG-based method, servers make decisions under minimal external information, and without any coordination with other servers.
%
%------------------------------------------------------->
\bibliographystyle{IEEEtran}
\bibliography{letter_ref}
%------------------------------------------------------->
\end{document}